\documentclass{article}
\usepackage{amssymb}
\usepackage{amsmath}
\usepackage{amsfonts}
\usepackage[mathcal]{eucal}
\usepackage{amscd}
\usepackage{xypic}
\newcommand{\A}{`}
\newcommand{\ks}{Kustaanheimo-Stiefel }
\newcommand{\g}{\mathbf{\Gamma}}

\newcommand{\lie}[1]{\boldsymbol{\CMcal{L}}_{#1}}
\newcommand{\de}{\mathrm{d}}
\newcommand{\rt}{\mathbb{R}^{3}_{\scriptscriptstyle{0}}}
\newcommand{\rqq}{\mathbb{R}^{4}_{\scriptscriptstyle{0}}}
\newcommand{\trt}{\mathrm{T}\mathbb{R}^{3}_{\scriptscriptstyle{0}}}
\newcommand{\trq}{\mathrm{T}\mathbb{R}^{4}_{\scriptscriptstyle{0}}}
\newcommand{\di}[1]{\cfrac{\partial}{\partial #1}}
\newcommand{\pks}{\CMcal{KS}}
\newcommand{\EL}{\mathcal{E}_{\mathcal{L}}}

\begin{document}
%
%
\title{Reduction and unfolding: the Kepler problem}%
\author{Antonella D'Avanzo and Giuseppe Marmo\\Dipartimento di Scienze Fisiche, Universit\'a Federico II\\
and\\
Istituto Nazionale di Fisica Nucleare - Sezione di Napoli\\
antonella.davanzo@na.infn.it, marmo@na.infn.it}
\date{}
\maketitle
%
%
\begin{abstract}
  In this paper we show, in a systematic way, how to relate the
  Kepler problem to the isotropic harmonic oscillator. Unlike
  previous approaches, our constructions are carried over in the
  Lagrangian formalism dealing, with second order vector fields. We
  therefore provide a tangent bundle version of the \ks map.
\end{abstract}
\numberwithin{equation}{section}
\section{Introduction}
Reduction procedures, providing a way to link a given dynamical
system with one on a lower dimensional manifold, have been
extensively  studied as an help in integrating the dynamics
(\cite{whitt}, \cite{MW74}, \cite{AM}, \cite{MSSV}, \cite{red},
\cite{red1}). It has been noted, however, that often a sort of
inverse of reduction, an \A\A unfolding'' procedure, can be more
effective for this purpose(\cite{KKS}). It has been shown
(\cite{pere}) that different classes of completely integrable
systems arise as reduction of free or \A\A simple'' ones with
higher degrees of freedom: in this approach, one can obtain the
solutions of the given dynamical system as a projection of the
ones of the higher dimensional system (exactly known). The
projection method would be therefore of interest in the problem of
integrating the dynamics to develope an unfolding tecnique. It is
clear that there is no unambiguous way to recognize that a given
system can be obtained as reduction of a free one, since there are
arbitrary elements in the reduction procedure itself (essentially
the choices of the submanifold of the carrier space invariant for
the dynamical system and of  the equivalence relation on it, cfr.
\cite{red}, \cite{red1}, \cite{genred}). Taking this into account,
one can try to focus on the main aspects of a dynamical system, so
to impose some conditions that the related system has to satisfy
and in this way reduce ambiguity. In this paper we develope such
an approach for the unfolding of the Kepler problem in three
degrees of freedom. This system has been widely studied (
\cite{SMO}, \cite{musto}, \cite{souriau}, \cite{mosercompl},
\cite{gy}, \cite{milnor}) and its relation with the isotropic
harmonic oscillator in four dimensions through the
Kustaanheimo-Stiefel map is well known \cite{ks}. We recover this
relation in a more general way,
this may be possibly used for the unfolding of a general system.\\
Here we only consider the Kepler problem in three degrees of
freedom, but it seems to us that our procedure can be fruitful
also in higher dimensions (for a different approach see
\cite{cordani1}, \cite{cordani2}).\\
The paper is organized as follows.\\
In section 2 we briefly recall the main elements of a reduction
procedure, giving two examples of completely integrable system
arising as reduction of a free system.\\
In section 3, after recalling few elements of Lagrangian
formalism, we point out the main aspects of the Kepler problem in
three dimensions, since they will be relevant in our unfolding
procedure; moreover, we introduce some properties of
reparametrized vector fields.\\
In section 4 we recover, from general consideration, the \ks
map as the main tool in our unfolding procedure.\\
In section 5 we arrive at an unfolding system for the Kepler
problem, and recognize that it coincides, up to reparametrization,
with a family of harmonic oscillators.\\
In section 6 we characterize the symmetry of the unfolding system
and find out the subalgebra of the constants of the motion for the
Kepler system.
\\ 
\numberwithin{equation}{section}\section{Reduction
procedure}\label{sec:red} In this section we point out the main
characteristics of a reduction procedure for second order systems,
since a thorough understanding of reduction procedure is essential
in performing its \A\A converse'', the unfolding procedure. In
particular, we treat two cases of nonlinear (completely
integrable) systems arising as reduction of free ones in higher
dimensions, in order to provide some motivations, through explicit
examples, for an
unfolding procedure.\\
\quad \\
Let $\g$ be a dynamical system on a carrier space $\CMcal{M}$, a
smooth manifold, namely a vector field in $\chi(\CMcal{M})$ (we
don't consider additional structures). Any reduction procedures
requires two ingredients:
\begin{itemize}
  \item[i)] a submanifold $\Sigma$ of $\CMcal{M}$ which is
  invariant for $\g$, that is to which $\g$ is tangent:
  \begin{equation}
\qquad    \g(m)\in \mathrm{T}_{m}\Sigma\qquad \forall m \in \Sigma
  \end{equation}
  \item[ii)]an equivalence relation $\approx$ on $\Sigma$ which is
  compatible with $\g$, that is (calling $\Phi^{t}_{\g}$ the flow of
  $\g$):
  \begin{equation}
  \qquad     m\approx m' \Rightarrow \Phi^{t}_{\g}(m)\approx
    \Phi^{t}_{\g}(m') \qquad \forall m,m'\in \Sigma
  \end{equation}
\end{itemize}
The reduced carrier space is the quotient manifold
$\tilde{\Sigma}=\Sigma/\approx $ and the reduced dynamical system
$\tilde{\g}$ is the projection of $\g_{\Sigma}$ (i.e. $\g$
restricted to $\Sigma$) on $\tilde{\Sigma}$ with respect to the
natural projection $\pi:\Sigma \rightarrow \Sigma/\approx\,\vdash
\, m \rightarrow \tilde{m}=[m]_{\approx}$ (well
defined thanks to the compatibility condition).\\
We recall that a vector field $\tilde{\g}\in\chi(\tilde{\Sigma})$
 is said to be the projection of $\g_{\Sigma}\in\chi(\Sigma)$ with respect
 to $\pi$, or that $\g_{\Sigma}$ is projectable onto $\tilde{\g}$ with
 respect to $\pi$ if the following diagram is commutative:
 \begin{equation}\label{eq:proiet1}
\xymatrix{T\Sigma \ar[r]^{T\pi}     & T\tilde{\Sigma}\\
 \Sigma \ar[u]^{\g_{\Sigma}} \ar[r]_{\pi}        & \tilde{\Sigma}
 \ar[u]_{\tilde{\g}}}
\end{equation}
\quad \\
Following this line, we give two examples of reduction of free
system that gives rise to nonlinear completely integrable systems;
for further details see \cite{MSSV}, \cite{genred}, \cite{red},
\cite{red1}.
\subsection{Example: radial reduction} Let us consider a free
system in 3 dimensions, described by the following vector field on
$\mathrm{T}\mathbb{R}^{3}$:
  \begin{equation}
    \g =v^{i} \frac{\partial}{\partial x^{i}} \qquad i=1 \ldots 3
  \end{equation}
corresponding to the following equation of the motion:
  \begin{equation}
    \ddot{\mathbf{r}}=0
  \end{equation}
where $\mathbf{r}$ is a vector in $\mathbb{R}^{3}$.\\
We will obtain a submanifold $\Sigma$ in $T\mathbb{R}^{3}$ fixing
the value of the energy, $\Sigma$ is invariant since $E$ is a
costant of the motion; the equivalence relation on $\Sigma$ will
be given in terms of the group $SO(3)$ of symmetry for the
system.\\
Introducing polar coordinates $ \mathbf{r}=r \hat{\mathbf{r}}$,
one has
  \begin{equation}\label{eq:coordpol}
    \dot{\mathbf{r}}=\dot{r} \hat{\mathbf{r}}+ r
    \dot{\Hat{\mathbf{r}}} ;\qquad
     \ddot{\mathbf{r}}=\ddot{r}
     \hat{\mathbf{r}}+2\dot{r}\dot{\Hat{\mathbf{r}}}+r\ddot{\Hat{\mathbf{r}}}
  \end{equation}
and, using the vectorial identities:
  \begin{equation}
    \Hat{\mathbf{r}} \cdot \Hat{\mathbf{r}}=1; \qquad \Hat{\mathbf{r}}
    \cdot \dot{\Hat{\mathbf{r}}}=0; \qquad
    (\dot{\Hat{\mathbf{r}}})^{2}=-\Hat{\mathbf{r}}\cdot
    \ddot{\Hat{\mathbf{r}}}
  \end{equation}
  and the equations of the motion, one gets:
  \begin{equation}\label{eq:ddotr}
    \ddot{r}=-r\Hat{\mathbf{r}}\ddot{\Hat{\mathbf{r}}}=r(\dot{\Hat{\mathbf{r}}})^{2}
  \end{equation}
  From the first equation \ref{eq:coordpol} it follows that:
  \begin{equation}
    \frac{\dot{\mathbf{r}}}{r}=\frac{\dot{r}}{r}\hat{\mathbf{r}}+
    \dot{\Hat{\mathbf{r}}} \; \Longrightarrow \;
     \left( \frac{\dot{\mathbf{r}}}{r}\right)^{2}=\left(\frac{\dot{r}}{r}\right)^{2}+
    (\dot{\Hat{\mathbf{r}}})^{2}
  \end{equation}
  so Eq. \ref{eq:ddotr} becomes:
  \begin{equation}
    \ddot{r}=\frac{(\dot{\mathbf{r}})^{2}}{r}-\frac{(\dot{r})^{2}}{r}
  \end{equation}
  For a free particle $2E=(\dot{\mathbf{r}})^{2}$; on the
  submanifolds of fixed energies
  \begin{equation}
    \Sigma_{E}=\{(\mathbf{r}\, , \mathbf{v})\in T\mathbb{R}^{3} \vdash
    (\mathbf{v})^{2}=2E=cost \}
  \end{equation}
  one obtains:
  \begin{equation}\label{eq:cristr}
    \ddot{r}=\frac{2E}{r}-\frac{(\dot{r})^{2}}{r}
  \end{equation}
  Since the equation defining $\Sigma_{E}$ is invariant under the action of $SO(3)$,
  the action of this group on
  $T\mathbb{R}^{3}$ can be restricted to an action on $\Sigma_{E}$.
  This provides an equivalence relation $\thickapprox_{\scriptscriptstyle{SO(3)}}$
  on  $\Sigma_{E}$: two points of $\Sigma_{E}$ are equivalent if they are connected
  by a transformation of $SO(3)$. So we obtain the quotient manifold:
  \begin{equation}
    \tilde{\Sigma}_{E}=\Sigma_{E}/SO(3)
  \end{equation}
The vector field $\g$ restricted to $\Sigma_{E}$ is projectable
since Eq. (\ref{eq:cristr}) is rotationally invariant; one has:
  \begin{equation}
    \tilde{\g}= v_{r} \frac{\partial}{\partial r} +\left(
    \frac{2E}{r}-\frac{(\dot{r})^{2}}{r}\right)\frac{\partial}{\partial v_{r}}
  \end{equation}
\textit{Remark}: if one considers $\Sigma_{\ell}$ the submanifolds
of fixed value for the angular momentum $\ell$, from
$\ell^{2}=(\dot{\mathbf{r}})^{2}(\mathbf{r})^{2}-(\mathbf{r}\cdot\dot{\mathbf{r}})$
one obtains:
\begin{equation}
  \tilde{\g}=v_{r}\di{r}+\frac{\ell^{2}}{r^{3}}\di{v_{r}}
\end{equation}
%
%
\subsection{Example: the Calogero-Moser system}
We consider again a reduction of the free motion in
$\mathbb{R}^{3}$; if we parametrize the elements of
$\mathbb{R}^{3}$ using $2\times 2$ symmetric matrices:
\begin{equation}
  X=\left(%
\begin{array}{cc}
  x_{1} & \frac{1}{\sqrt{2}}x_{2} \\
  \frac{1}{\sqrt{2}}x_{2} & x_{3} \\
\end{array}%
\right)
\end{equation}
the equations of motion for the free particle in 3 dimensions can
be written as:
\begin{equation}
  \ddot{X}=0
\end{equation}
and the Lagrangian function becomes:
\begin{equation}
  \mathcal{L}_{0}=\frac{1}{2}\mathrm{Tr}\dot{X}^{2}=\frac{1}{2}(\dot{x}^{2}_{1}
  +\dot{x}^{2}_{2}+\dot{x}^{2}_{3})
\end{equation}
The matrix:
\begin{equation}
  M=[X,\dot{X}]
\end{equation}
is a constant of the motion; being antisymmetric, it can be
written as:
\begin{equation}
  M=\ell\sigma, \qquad \sigma =\left(%
\begin{array}{cc}
  0 & 1 \\
  -1 & 0 \\
\end{array}%
\right)
\end{equation}
where $\ell$ is the modulus of the angular momentum.\\
Since $X$ is a real symmetric matrix, it can be diagonalized by
elements of the rotation group $SO(2)$:
\begin{equation}
  X=G\,Q\,G^{-1}
\end{equation}
where:
\begin{equation}
  Q=\left(%
\begin{array}{cc}
  q_{1} & 0 \\
  0 & q_{2} \\
\end{array}%
\right),\qquad G=\left(%
\begin{array}{cc}
  \cos\varphi & \sin\varphi \\
  -\sin\varphi & \cos\varphi \\
\end{array}%
\right)
\end{equation}
From this one obtains:
\begin{equation}\label{eq:X}
  \dot{X}=[\dot{G}\,G^{-1},\,X]+G\,\dot{Q}\,G^{-1}=G\,\left([\,G^{-1}\,\dot{G}\,,Q]+\dot{Q}\right)\,G^{-1}
\end{equation}
and:
\begin{equation}
 M=[X,[\dot{G}\,G^{-1},X\,]]=G\,[Q,\,[G^{-1}\,\dot{G},\,Q\,]]\,G^{-1}
\end{equation}
Moreover:
\begin{equation}
  G^{-1}\,\dot{G}=\dot{G}\,G^{-1}=\dot{\varphi}\sigma
\end{equation}
Using these relations one can evaluate the trace of $M\sigma$ to
find the value of $\ell$:
\begin{equation}\label{eq:phi}
\ell=-\frac{1}{2}\mathrm{Tr}(M\sigma)=\dot{\varphi}(q_{2}-q_{1})^{2}
\end{equation}
from which one can obtain $\dot{\varphi}$. Deriving \ref{eq:X}
with respect to time, one gets:
\begin{equation}
  \ddot{X}=\dot{\varphi}^{2}\,[\,\sigma,\,[\,\sigma,\,Q\,]]
\end{equation}
that is, substituting $\dot{\varphi}$ from Eq.~\ref{eq:phi}:
\begin
{gather}
  \ddot{q}_{1}=-\frac{2\ell^{2}}{(q_{2}-q_{1})^{3}} \nonumber\\
  \ddot{q}_{2}=\frac{2\ell^{2}}{(q_{2}-q_{1})^{3}}
\end{gather}
Since $\ell$ is a constant of the motion, the condition
$\ell=const$ provides invariant submanifolds $\Sigma_{\ell}$ in
the tangent bundle of the space of symmetric matrices; since this
space covers $\mathrm{T}\mathbb{R}^{2}$, one can project from each
one of these submanifolds and find a family of dynamical systems
in the variables $q_{1}, q_{2}$, that is a second order dynamical
system on $\mathrm{T}\mathbb{R}^{2}$. When $\ell=0$ we get the
free particle in 2 dimensions; when $\ell\neq 0$ we get the
Calogero-Moser system in $\mathbb{R}^{2}$ (\cite{moser},
\cite{calogero},\cite{pere}).
 \vspace{8mm}\\
 It is clear from these examples that having reduced the dimensions of the dynamical
 system does not always provide a simpler system: we have obtained
nonlinear systems starting with a linear one; moreover, in the
first case the integration is not immediate. Such cases suggest
that it could be reasonable for the integration of the dynamics to
investigate if a given nonlinear system could arise as reduction
of a linear one, so to obtain the solutions of the former
projecting
those of the latter.\\
We shall try now to apply an unfolding procedure to the
three-dimensional Kepler problem.
\numberwithin{equation}{section}\section{The Kepler problem} Our
first goal is to enumerate the main aspects of the Kepler system,
in order to understand what properties one requires for
the unfolding system.\\
Before doing this we recall some elements
of the geometry of Lagrangian formalism on the the tangent bundle,
assumed to be the carrier space for the dynamics.
\subsection{Few elements of Lagrangian
formalism}\label{sec:lagrform}
It is well known that it is possible to formulate the Lagrangian
formalism in an intrinsic, coordinate free version, involving
objects characteristic of the geometry of the tangent bundle; this
is what we briefly recall here, referring to \cite{lagr}
for details.\\
On a tangent bundle there are two natural tensor fields, they
essentially characterize its structure: the \emph{vertical
endomorphism} $\textrm{S}$ and the \emph{dilation vector field}
$\Delta$, given in natural coordinates by: \begin{equation}
S=\de
x^{i} \otimes \di{v^{i}}; \qquad \Delta=v^{i}\di{v^{i}}
\end{equation}
In terms of these objects, a \emph{second order vector field} $\g$
can be defined in intrinsec terms by:
\begin{equation}
  S(\g)=\Delta
\end{equation}
Moreover, it is possible to associate to $S$ a \emph{ generalized
derivation} $\de_{S}f=\de f \circ S$, in coordinates $\de_{S}
f=\cfrac{\partial
f}{\partial v^{i}} \de x^{i}$.\\
Given a Lagrangian function $\mathcal{L}$, one can define the
\emph{Cartan 1-form} of $\mathcal{L}$:
\begin{equation}
 \theta_{\mathcal{L}}:=\de_{S}\mathcal{L}
\end{equation}
and the \emph{Cartan 2-form}:
\begin{equation}
  \omega_{\mathcal{L}}=-\de \theta_{\mathcal{L}}
\end{equation}
If $\omega_{\mathcal{L}}$ is non degenerate, i.e. symplectic,
$\mathcal{L}$ is said
to be regular.\\
Having introduced these objects, it is possible to write the
Euler-Lagrange equation in the following coordinate-free way:
\begin{equation}\label{eq:eullagr}
  \lie{\g}\theta_{\mathcal{L}}-\de \mathcal{L}=0
\end{equation}
where $\g$ is a vector field to be determined. When $\mathcal{L}$
is regular, $\g$ is second order: in this case, after introducing
the \emph{energy function}
\begin{equation}\label{eq:ef}
  \mathcal{E}_{\mathcal{L}}:=\lie{\Delta}\mathcal{L}-\mathcal{L}
\end{equation}
one can rewrite equation \ref{eq:eullagr} in the following
equivalent way:
\begin{equation}
  i_{\g}\omega_{\mathcal{L}}=\de \mathcal{E}_{\mathcal{L}}
\end{equation}
So, when the Lagrangian function is regular, we have obtained a
symplectic formulation of the Lagrangian dynamics directly on the
tangent bundle. It is now possible to define Poisson brackets on
the tangent bundle as one usually does on any symplectic manifold:
\begin{equation}\label{eq:pbl}
  \{f,g\}_{\mathcal{L}}:=\omega(\mathbf{X}_{f},\mathbf{X}_{g})
\end{equation}
where the vector fields are obtained solving
$i_{\mathbf{X}_{f}}\omega_{\mathcal{L}}=\de f$ and
$i_{\mathbf{X}_{g}}\omega_{\mathcal{L}}=\de g$.\\
In this way the vector space of  functions on the tangent bundle
acquires the structure of a Lie algebra under  Poisson bracket.
\\We note that now the symplectic 2-form
depends on $\mathcal{L}$ and so does the Poisson bracket (see \cite{Fpr}).\\

\subsection{The Kepler system}
By Kepler system we mean the model of a point particle subjected
to a central force proportional to the inverse of the square of
the radius:
\begin{equation}
 \mathbf{F}_{K}=-\frac{k\mathbf{x}}{r^{3}}
\end{equation}
The configuration space of the Kepler system is $\mathbb{R}^{3}$
minus the origin; in the following we will denote
$\mathbb{R}^{n}-\{ 0\}=\mathbb{R}^{n}_{0}$ for brevity.\\
To the equations of the motion, that is a second order system of
differential equations in $\rt$, it is possible to associate a
first order system on the tangent bundle of $\rt$, whose solutions
are the integral curves of the vector field\footnote{in the
following latins indices run from 1 to 3, greek ones from 0 to 3
and the summation over repeated indeces is adopted}:
\begin{equation}
  \g_{K}=v^{i}\di{x^{i}}-\frac{k x^{i}}{r^{3}}\di{v^{i}}
\end{equation}
The vector field $\g_{K}\in\chi(\trt)$ is a \emph{second order
vector field} (cfr. sec. \ref{sec:lagrform}); it defines a system
of second order differential equations on the configuration
space.\\
It is well known that the Kepler problem admits a Lagrangian
formulation, with Lagrangian function:
\begin{equation}
  \mathcal{L}_{K}=\frac{1}{2}v^{i}v^{i}+\frac{k}{r}
\end{equation}
where $r=\sqrt{x^{i}x^{i}}$ and we have set the mass $m=1$.\\
The energy associated to this Lagrangian is (cfr.
Eq.~(\ref{eq:ef})):
\begin{equation}
  \mathcal{E}_{\mathcal{L}_{K}}=\frac{1}{2}v^{i}v^{i}-\frac{k}{r}
\end{equation}
and the Cartan 1- and 2- forms are, respectively:
\begin{gather}
  \theta_{\mathcal{L}_{K}}=v^{i} \, \de x^{i}\\
  \omega_{\mathcal{L}_{K}}=-\de \theta_{\mathcal{L}_{K}} = \de
  x^{i} \wedge \de v^{i}
\end{gather}
such that one has:
\begin{equation}
  i_{\g_{K}}\omega_{\mathcal{L}_{K}}=\de \mathcal{E}_{\mathcal{L}}
\end{equation}
We notice that we can express all these objects in terms of the
Euclidean metric of $\mathbb{R}^{3}$.\\
The Poisson bracket obtained from the symplectic form
$\omega_{\mathcal{L}_{K}}$ as in Eq.~(\ref{eq:pbl}) is:
\begin{equation}
  \{f,g\}_{\mathcal{L}_{K}}=\frac{\partial f}{\partial x^{i}}\frac{\partial g}{\partial v^{i}}
  -\frac{\partial f}{\partial v^{i}}\frac{\partial g}{\partial x^{i}}
\end{equation}
It is well known that constants of the motions for the Kepler
problem are:
\begin{gather}
  L_{i}=\epsilon_{ijk}x^{i}v^{k}\\
  A_{i}=\left(\frac{k
  x^{i}}{r}+\epsilon_{ijk}v_{j}L_{k}\right)
\end{gather}
respectively the angular momentum vector and the Runge-Lenz
vector, satisfying the equation:
\begin{equation}\label{eq:LA}
 L_{i}A_{i}=0
\end{equation}
We recall that in the association between symmetries and constants
of the motion by means of the Lagrangian symplectic structure
$\omega_{\mathcal{L}}$ only the first one arise from point
transformations, while the second comes from a so called \A\A
extended symmetry'', i.e. a dynamical symmetry which does not
respect
the tangent bundle structure of the carrier space.\\
The commutation relations among the constants of the motion are
the following:
\begin{align}\label{eq:marl}
  \{L_{i},L_{j}\}_{\mathcal{L}_{K}}=\epsilon_{ijk}L_{k}; &\qquad
  \{A_{i},A_{j}\}_{\mathcal{L}_{K}}=-2\mathcal{E}_{\mathcal{L}}\epsilon_{ijk}L_{k} \nonumber\\
  \{L_{i},A_{j}\}_{\mathcal{L}_{K}}=\epsilon_{ijk}A_{k} & \,
\end{align}
Since the energy function $\mathcal{E}_{\mathcal{L}_{K}}$ belongs
to the center of the algebra of the constants of the motion, we
can rescale the Runge-Lenz vector by appropriate functions of the
energy (of constant sign). We can divide $\trt$ into three sets,
the two open regions when $\mathcal{E}_{\mathcal{L}_{K}}<0$, and
$\mathcal{E}_{\mathcal{L}_{K}}>0$ and their boundary
$\mathcal{E}_{\mathcal{L}_{K}}=0$: when
$\mathcal{E}_{\mathcal{L}_{K}}<0$, we can rescale $A^{i}$ by
$\sqrt{-2\mathcal{E}_{\mathcal{L}_{K}}}$, to get an
$\mathfrak{o}(4)$ algebra; where
$\mathcal{E}_{\mathcal{L}_{K}}>0$, we can rescale $A^{i}$ by
$\sqrt{2\mathcal{E}_{\mathcal{L}_{K}}}$, to get an
$\mathfrak{o}(3,1)$ algebra; when $\mathcal{E}_{\mathcal{L}_{K}}=0$
we get the Eucledean algebra in 3D.  \\
 The Kepler system is completely integrable and maximally
superintegrable. We recall that a system with $n$ degrees of
freedom is said to be maximally superintegrable when the largest
number of constants of the motion which are essentially
independent is $2n-1$;  a set of functions $f^{1},\ldots f^{k}\in
\CMcal{F}(\CMcal{M})$ is said to be essentially independent if the
set $\{x^{i}\in \CMcal{M}\vdash \de f^{1}(x)\wedge \ldots \wedge
\de f^{k}(x)=0\}$ has no interior points. The Kepler system has
2n-1=5 constants of the motion essentially independent, they are
the components of the angular momentum and of the Runge-Lenz
vector (only five of them are independent because of the
constraint
\ref{eq:LA}).\\ 
Another important feature of the Kepler system is
that it has closed orbits for an open portion of the phase space,
the one
corresponding to negative energies.\\
In the search for a system the Kepler problem is a reduction of,
we are suggested to impose the condition of closed orbits;
moreover, it seems reasonable to preserve also the property of
superintegrability. Along the line of section \ref{sec:red}, the
first attempt would be in terms of \emph{linear maximally
superintegrable systems with closed orbits}, and
the only system that satisfies all these conditions is the harmonic oscillator.\\
We have required that the systems from which one can obtain the
Kepler problem by reduction have to share some properties of the
Kepler problem, and in this way we have restricted the ambiguity
in
the unfolding procedure.\\
Few comments are in order at this point. First, there is an
obstruction to the correspondence between harmonic oscillator and
Kepler problem given by the energy-period theorem. It says that,
if the period $T$ of a periodic Hamiltonian system is a (at least)
$C^{1}$ function of the carrier space, then $\de H \wedge \de
T=0$; so the period is a function of the energy: $T=T(E)$ (see e.g. \cite{AM}).\\
If we have a system $\g$ which reduces to $\tilde{\g}$, as in
section \ref{sec:red}, from the equivariance of the flows (in the
notations of section \ref{sec:red}):
\begin{equation}
\Phi^{t}_{\tilde{\g}}\circ\pi=\pi\circ\Phi^{t}_{\g_{\Sigma}}
\end{equation}
it follows that the period $T$ of the unfolding system is also a
period for the reduced system:
\begin{equation}
  \Phi^{t+T}_{\g}(m)=\Phi^{t}_{\g}(m) \Rightarrow
  \Phi^{t+T}_{\tilde{\g}}(\tilde{m})=\Phi^{t}_{\tilde{\g}}(\tilde{m})
\end{equation}
Thus, an isocronous system cannot reduce to an non-isocronous
sytem, and vice versa.\\
While for the Kepler problem orbits with different energies do
have different periods, it is well known that the period of the
harmonic oscillator is constant, independent from the energy: from
the previous result, it follows that we shall not be able to map
the isotropic harmonic oscillator motions onto the motions of the
Kepler problem. So, if integral curves of the isotropic harmonic
oscillator project to integral curves of the Kepler system, to
match the periods we are obliged to use a one parameter family of
different oscillators whose period depends on the energy of the
Kepler system, and the correspondence can be done only on
submanifolds of fixed energy.\\
The second comment concerns another important difference between
the two systems: while the vector field of the harmonic oscillator
is complete, the one of the Kepler problem is not complete. So the
problem arises of how to link two such systems. But a well known
theorem (see e.g. \cite{godbillon},~Chap.V-1.8) asserts that,
given a vector field $\mathbf{X}$ on a paracompact manifold
$\CMcal{M}$, there exists a strictly positive function $f$ on
$\CMcal{M}$, of the same differentiability class as $\mathbf{X}$,
such that $\mathbf{\tilde{X}}=f\cdot\mathbf{X}$ is complete. Thus
we should expect that at a certain point in our unfolding
procedure there will be a need for a  reparametrization of the
vector field involved: therefore it is appropriate to point out
some aspects connected with reparametrization.
\subsection{Reparametrized  vector fields: some properties}\label{sec:reptb}
First of all, we notice that the integral
curves of $\mathbf{X}$ and $\mathbf{\tilde{X}}=f\cdot \mathbf{X}$,
are the same, but
their parametrization changes.\\
Anyway, the constants of the motion do not depend on the
parametrization, since:
\begin{equation}
  \lie{\mathbf{X}}h=0 \Rightarrow \lie{\mathbf{\tilde{X}}}h=f\lie{\mathbf{X}}h=0
\end{equation}
However symmetries, represented by vector fields, do depend on the
parametrization, since from $[\mathbf{X},\mathbf{Y}]=0$ one only
gets
$[f\mathbf{X},\mathbf{Y}]=(\lie{\mathbf{Y}}f)\mathbf{X} $ \\
Obviously, the reparametrization of a vector field does not
preserve some other properties. For intance, if $\mathbf{X}$ is
Hamiltonian, from $i_{\mathbf{X}}\omega=\de h$ it follows that
$i_{f\mathbf{X}}\omega=f\de h$,  so $\tilde{\mathbf{X}}$ is not
Hamiltonian unless $\de f\wedge\de h=0$.\\
Moreover, if $\mathbf{X}\in\chi(\mathrm{T}\mathcal{Q})$ is a
second order vector field, $\tilde{\mathbf{X}}$ in general will
not have the same property; in fact:
\begin{equation}
  S(\mathbf{X})=\Delta \Rightarrow S(f\mathbf{X})=f\Delta
\end{equation}
However, it is possible to endow $\mathrm{T}\mathcal{Q}=\CMcal{M}$
with a different structure of tangent bundle with respect to which
the reparametrized vector field is second order. We now sketch a
possible procedure,
referring to \cite{inprep} for details.\\
We have already noticed that two objects characterize the geometry
of the tangent bundle: the vertical endomorphism $S$ and the
dilation vector field $\Delta$. Actually, it is possible to prove
that, given a vector field and a 1-1 tensor field on a manifold
$\CMcal{M}$ verifying the same joint properties of $\Delta$ and
$S$, they characterize \emph{uniquely} the tangent bundle
structure on $\CMcal{M}$, that is a unique manifold $\mathcal{B}$
exists such that $\CMcal{M}\thicksim \mathrm{T}\mathcal{B}$ (see
\cite{teotang}
for details).\\
Using this result, given a vector field $\mathbf{X}$ on a manifold
$\CMcal{M}$, it is possible to provide $\CMcal{M}$ with a tangent
bundle structure with respect to which $\mathbf{X}$ is second order,
if some appropriate conditions are satisfied.\\
Let dim$\CMcal{M}=2$n, and let $g^{i}$ be n functionally
independent functions on $\CMcal{M}$, that is $\de g^{1} \wedge
\de g^{2}\wedge \ldots \wedge \de g^{n}\neq 0$; let us suppose
that n functions $f^{i}$ exist such that:
\begin{equation}\label{eq:hp}
  f^{i}=\lie{\mathbf{X}}g^{i}
\end{equation}
(possibly only locally) and that they satisfy the relation:
\begin{equation}
  \de g^{1} \wedge \ldots \wedge \de g^{n} \wedge \de f^{1}\wedge
  \ldots \de f^{n}\neq 0
\end{equation} The set $(f^{i}, g^{i}$) will provide a set of coordinates on
$\CMcal{M}$ (at least locally); so one can build a 1-1 tensor
field and a vector field:
\begin{equation}\label{eq:sdelta}
S:=\de g^{i}\otimes \di{f^{i}}; \qquad  \Delta:=f^{i}\di{f^{i}}
\end{equation}
that verify the joint properties we mentioned above. By virtue of
the theorem in \cite{teotang}, they define (at least locally) on
$\CMcal{M}$ a structure of tangent bundle, where the $g^{i}$ can
be considered the base coordinates and the $f^{i}$ the fiber
coordinates\footnote{we notice that the $g^{i}$ are an algebra
with respect to the pointwise product, while the $f^{i}$ are a
module space}. The vector field $\mathbf{X}$ turns out to be
second order with respect to this structures, since:
\begin{equation}
  S(\mathbf{X})=\Delta
\end{equation}
using Eq. \ref{eq:hp}, that was our main assumption.\\
If equation \ref{eq:hp} is verified locally, one has to repeat
this construction for each open set where Eq. \ref{eq:hp} is
verified and take into account transition functions which should
behave as ``point transformations'', i.e. tangent bundle isomorphisms.\\
Incidentally, we notice that the same manifold $\CMcal{M}$ can
acquire different tangent bundle structures. For instance, one can
multiply the same set of functions $g^{i}$ by a function of some
constants of the motion $C^{\alpha}$ for $\mathbf{X}$ (that is
$\lie{\mathbf{X}}C^{\alpha}=0$); so one has:
\begin{equation}
G^{i}=h(C^{\alpha})g^{i}\Rightarrow
\lie{\mathbf{X}}G^{i}=h(C^{\alpha})g^{i}:=F^{i}
\end{equation}
so the functions $(G^{i},F^{i})$ defines different $S'$ and
$\Delta'$, and so a different structure of tangent bundle on
$\CMcal{M}$.
%
\numberwithin{equation}{section}\section{The  \ks map} So far,
from general considerations, we have considerably reduced the
ambiguity of the unfolding procedure, having identified as a
candidate unfolding system only (a family of) harmonic
oscillators. Of course, we still have ambiguity on its degrees of
freedom: the minimal condition we can impose is that they have to
be stricly greater then those of the Kepler problem, since the
latter has to arise as a reduction of the former. So, calling $n$
this number of degrees of freedom, we have $n>3$. For simplicity
one can consider first the cases with lowest $n$, and investigate
if the correspondence can be realized for, say, $n=4$. So one has
to look for a map $ \trq \rightarrow \trt$. Since the angular
momentum for the Kepler problem is a constant of the motion
associated to point transformations, and since constants of the
motion of the reduced sistem are projection of the unfolding ones,
we can ask that the reduction procedure respect this structure,
that is it comes from a map between the configuration spaces $\pi:
\rqq \rightarrow \rt$. Since we have to perform a reduction, we
look for a covering of $\rt$ with $\rqq$. Because
$\rt=S^{2}\times\mathbb{R}^{+}$ and
$\rqq=S^{3}\times\mathbb{R}^{+}$ we may start from a covering map
$\pi_{H}:S^{3} \rightarrow S^{2}$ and extend it. Identifying
$S^{3}$ with $SU(2)$, we can represent it in terms of matrices:
\begin{align}
s&=\left(%
\begin{array}{cc}
  y_{1}+iy_{2} & y_{0}+iy_{3} \\
  -y_{0}+iy_{3} & y_{1}-iy_{2} \\
\end{array}
\right) &;\qquad & y_{\alpha} \vdash \sum_{\alpha=0}^{3}
y_{\alpha}y_{\alpha} = \mathrm{det}(g) = 1.
\end{align}
and so define the Hopf map (\cite{hopfmap}) by setting:
\begin{equation}
  \pks: s \rightarrow \vec{x} \vdash s \sigma_{3} s^{-1}=x^{i}\sigma_{i}
\end{equation}
where $\sigma_{i}$ are the Pauli matrices (and
$\sigma_{0}=\mathbb{I}$).\\
There are several ways to extend the Hopf map to $\rqq \rightarrow
\rt$. A natural one is obtained by introducing polar coordinates
in $\rqq=S^{3}\times \mathbb{R}^{+}$ and setting:
\begin{equation}
 \; \qquad g=Rs    \;\; \mathrm{with}\;\; s\in SU(2),\, R\in \mathbb{R}^{+}
\end{equation}
and, recalling that $s^{-1}=s^{\dagger}$ for $s \in SU(2)$, we
define:
\begin{equation}\label{eq:def}
  \pks: g \rightarrow \vec{x}\vdash \; x^{k}\sigma_{k}= g \sigma_{3}
  g^{\dagger}=R^{2}s \sigma_{3} s^{-1}
\end{equation}
or, alternatively:
\begin{equation}
 \begin{array}{lll}
    x_{1}&=& 2(y_{1}y_{3}+y_{2}y_{0})\\
    x_{2}&=& 2(y_{2}y_{3}-y_{1}y_{0})\\
    x_{3}&=& y_{1}^{2}+y_{2}^{2}-y_{3}^{2}-y_{4}^{2}.
\end{array}
\end{equation}
From $\mathrm{det}g=R$, we find:
\begin{equation}
  -x^{i}x^{i}=\mathrm{det}(g\sigma_{3} g^{\dagger})=-\mathrm{det}(g^{\dagger}g)=-R^{4}
\end{equation}
or:
\begin{equation}\label{eq:Rr}
  r=R^{2}
\end{equation}
From Eq.~(\ref{eq:def}) it follows that the one parameter group
$\exp(i \lambda \sigma_{3})$ defines the fibers of the fibration
$U(1)\rightarrow \rqq \rightarrow \rt$; we notice that this group
acts by right moltiplication.\\
This map extends naturally to $\trq \rightarrow \trt$:
\begin{equation}
\begin{array}{ll}
\mathrm{T}(\pks): &
  \begin{array}{lll}
   v_{1}&=& 2(y_{1}u_{3}+y_{2}u_{0}+y_{3}u_{1}+y_{0}u_{2})\\
   v_{2}&=& 2(y_{3}u_{2}+y_{2}u_{3}-y_{1}u_{0}-y_{0}u_{1})\\
   v_{3}&=& y_{1}u_{1}+y_{2}u_{2}-y_{3}u_{3}-y_{0}u_{0}
\end{array}
\end{array}
\end{equation}
and one has the fibration $\mathrm{T}U(1)\rightarrow \trq \rightarrow \trt$.\\
In this way we have recovered the \ks map, introduced in
\cite{ks}, where the authors extablished a relation between the
solutions of the isotropic harmonic oscillator and those of the
Kepler problem. Here we have tried to arrive at this map in a
constructive way, as far as it has been possible: we have made
some general requirements, but also arbitrary choices, for
instance the number of dimensions and  the extension of  the Hopf
map; anyway, all our considerations relied on general properties
of the system, so  could be hopefully
repeated for other systems, with the appropriate changes.\\
%
%
%
\numberwithin{equation}{section}\section{Unfolding of the Kepler
problem}\label{sec:unfolding}
Having introduced all the elements we need, we can now look for
the motions in 4 dimensions which may be related to Keplerian
motions in 3 dimensions. At this point   the  Lagrangian character
of the Kepler problem may be put to work. Since
$i_{\g_{K}}\omega_{\mathcal{L}_{K}}=\de
\mathcal{E}_{\mathcal{L}_{K}}$, pulling back
$\omega_{\mathcal{L}_{K}}$ and $\mathcal{E}_{\mathcal{L}_{K}}$
with $\mathrm{T}\pks$, after appropriate considerations, we can
obtain a symplectic vector field whose
projection is the Kepler vector field.\\
As for the pull back of $\omega_{\mathcal{L}_{K}}$ and
$\mathcal{E}_{\mathcal{L}_{K}}$ we may instead consider the pull
back of the Euclidean metric $\de x^{i} \otimes \de x^{i}$, from
which $\mathcal{L}_{K}$ and $E_{\mathcal{L}_{K}}$ are derived:
\begin{align}
  (\mathrm{T}\pks)^{*}(\de x^{i} \otimes \de
  x^{i}) &=(\mathrm{T}\pks)^{*}\left(\frac{1}{2}\mathrm{Tr}[(\de x^{k}\sigma_{k})\otimes (\de
  x^{k}\sigma_{k})]\right)= \nonumber\\
=&\frac{1}{2}\mathrm{Tr}[\de(g \sigma_{3}g^{\dagger})\otimes\de(g
\sigma_{3}g^{\dagger})]
\end{align}
Using that $g=Rs$, one gets:
\begin{equation}
  \de(g\sigma_{3}g^{\dagger})=\de R^{2}\,s\sigma_{3}
  s^{-1}+R^{2}\de(s\sigma_{3} s^{-1})
\end{equation}
so:
\begin{gather}
  \frac{1}{2}\mathrm{Tr}(\de R^{2}\,s\sigma_{3}
  s^{-1}+R^{2}\de(s\sigma_{3} s^{-1}))\otimes(\de R^{2}\,s\sigma_{3}
  s^{-1}+R^{2}\de(s\sigma_{3} s^{-1}))=\nonumber\\
  \frac{1}{2}[\de R^{2}\otimes\de
  R^{2}\mathrm{Tr}(s\sigma_{3}s^{-1}s\sigma_{3}s^{-1})
  +R^{4}\mathrm{Tr}[\de(s\sigma_{3}s^{-1})\otimes\de(s\sigma_{3}s^{-1})]\nonumber\\
  +R^{2}\de
  R^{2}\otimes\mathrm{Tr}[(s\sigma_{3}s^{-1})\de(s\sigma_{3}s^{-1})]+
   R^{2}\mathrm{Tr}[(s\sigma_{3}s^{-1})\de(s\sigma_{3}s^{-1})]\otimes\de
  R^{2}]
\end{gather}
The trace in the first term is 1, the last two terms are zero
because the trace involved is zero. As for the second term, using
that:
\begin{gather}
\de(s\sigma_{3}s^{-1})=\de(s)\sigma_{3}s^{-1}+s\sigma_{3}\de
(s^{-1})=\de s \,\sigma_{3}s^{-1}-s\sigma_{3}s^{-1}\de s\,
s^{-1}=\nonumber\\
=[s^{-1}\de s,
\sigma_{3}]=i[\sigma_{k}\theta^{k},\sigma_{3}]=2i\epsilon_{k3i}\sigma_{i}\theta^{k}
\end{gather}
where the square bracket stands for the commutator, and we have
used the relation $s^{-1}\de s=i \sigma_{k}\theta^{k}$, where
$\theta^{k}$ are the left invariant one-forms of
$\chi^{*}(SU(2))$. So we get:
\begin{gather}
\mathrm{Tr}[\de(s\sigma_{3}s^{-1})\otimes\de(s\sigma_{3}s^{-1})]=\theta^{1}\otimes
\theta^{1}+\theta^{2}\otimes\theta^{2}
\end{gather}
As for the pull back of the metric, one finally obtains:
\begin{equation}\label{eq:pullbmetr}
  (\mathrm{T}\pks)^{*}(\de x^{i} \otimes \de
  x^{i})=4R^{2} [\de R \otimes \de R
  +R^{2}(\theta_{1}\otimes\theta_{1}+\theta_{2}\otimes\theta_{2})]
\end{equation}
Of course, it is a degenerate quadratic form; but a comparison
with the conformally flat metric:
\begin{equation}\label{eq:matricac}
  g_{C}=4R^{2} [\de R \otimes \de R
  +R^{2}(\theta_{1}\otimes\theta_{1}+\theta_{2}\otimes\theta_{2}+\theta_{3}\otimes\theta_{3})]
\end{equation}
suggests that we complete it by adding the missing term
$\theta_{3}\otimes\theta_{3}$. We notice, in fact, that it doesn't
affect the kinetic energy in $\trt$, since if one has\footnote{in
the following $\dot{\theta}_{k}=i_{\g}\tau^{*}_{\rqq}\theta_{k}$}
in $\trq$:
\begin{equation}
  T=4R^{2}[\dot{R}^{2}+R^{2}(\dot{\theta}_{1}^{2}+\dot{\theta}_{2}^{2}+\dot{\theta}_{3}^{2})]
\end{equation}
the last term, the added one, is invariant under the right
multiplication by $\exp(i\lambda\sigma_{3})$, the group with
respect to which one has to perform the reduction; so in the
reduction procedure, choosing $R^{4}\dot{\theta}_{3}=0$ as the
invariant manifold $\Sigma_{0}$ (see section \ref{sec:red})  gives
the Kepler system (we will show later that it actually is
invariant). Incidentally, we note that $R^{4}\dot{\theta}_{3}$ is
the Hamiltonian function for the tangent lift of the former $U(1)$
action.
\\ With these considerations,
after the pull-back we get a \A\A conformal Kepler problem'' in
$\rqq$, described by the Lagrangian:
\begin{equation}
  \mathcal{L}=2R^{2}[\dot{R}^{2}+R^{2}(\dot{\theta}_{1}^{2}
  +\dot{\theta}_{2}^{2}+\dot{\theta}_{3}^{2})]+\frac{1}{R^{2}}
\end{equation}
to which is associated the energy:
\begin{equation}
  \EL=2R^{2}[\dot{R}^{2}+R^{2}(\dot{\theta}_{1}^{2}
  +\dot{\theta}_{2}^{2}+\dot{\theta}_{3}^{2})]-\frac{1}{R^{2}}
\end{equation}
and the Cartan 1-form and symplectic 2-form are given by:
\begin{gather}
  \theta_{\mathcal{L}}=4 R^{2}(\dot{R}\,\de R+R^{2}\,\dot{\theta}_{k}\,\theta^{k})\nonumber\\
  \omega_{\mathcal{L}}= -4\,R^{2}\,\de\dot{R}\wedge \de R
  -16\,R^{3}\,\dot{\theta}_{k}\,\,\de\,R\wedge
  \theta^{k}  -4\,R^{4}(\de\dot{\theta}_{k}\wedge\theta^{k}+\dot{\theta}_{k}\de\theta^{k})
\end{gather}
For future convenience we express all these objects in cartesian
coordinates $(y^{\alpha}, \,u^{\alpha})$:
\begin{align}\label{eq:objects}
  \qquad \mathcal{L}&=4R^{2}\frac{1}{2}u^{i}u^{i}+\frac{k}{R^{2}}; &
   \qquad \EL&=4R^{2}\frac{1}{2}u^{i}u^{i}-\frac{k}{R^{2}}\\
  \qquad \theta_{\mathcal{L}}&=4R^{2}u^{i} \de y^{i} &
  \qquad \omega_{\mathcal{L}}&=-8y^{\beta}u^{\alpha}\de y^{\beta}\wedge
  \de y^{\alpha}-4R^{2}\de u^{\alpha}\wedge \de y^{\alpha}
  \nonumber
\end{align}
The dynamical vector field obtained from
$i_{\g}\omega_{\mathcal{L}}=\de \EL$ is:
\begin{equation}
\begin{array}{lll}
 \g &=& u^{\alpha}\di{y^{\alpha}}+F^{\alpha}\di{u^{\alpha}}=\\
  \quad &=&u^{\alpha}\di{y^{\alpha}}+
  \left(\cfrac{u^{2}}{R^{2}}y^{\alpha}-\cfrac{k}{2R^{6}}y^{\alpha}
  -2\cfrac{u^{\beta}y^{\beta}}{R^{2}}u^{\alpha}\right)\di{u^{\alpha}}
  \end{array}
 \end{equation}
At this point, we have completely determined a dynamical system in
$\rqq$ that  gives the Kepler system by reduction. More precisely,
one first has to restrict to the submanifold $\Sigma_{0}$ (see
sec. \ref{sec:red}) where $R^{4}\dot{\theta}_{3}=0$, which is
invariant for $\g$ since $\lie{\g}(R^{4}\dot{\theta}_{3})=0$. On
this submanifold, one chooses the equivalence relation defined by
the action of the group $U(1)$, tangent lift of the right
multiplication by $\exp(i\lambda\sigma_{3})$, i.e. two points are
equivalent if they belong to the same orbit of the group. This
equivalence relation is compatible with $\g$, because the group
under consideration is a simmetry group for $\g$, since its
Hamiltonian $R^{4}\dot{\theta}_{3}$ is a constant of the motion.
The reduced system obtained in this way is the Kepler system on
$\trt$.
\\Moreover, one can consider the algebra of the constants of the
motion of the conformal Kepler problem, where the Poisson bracket
is obtained from the symplectic 2-form as previously shown: it is
possible to obtain the constants of the motion of the Kepler
problem as a subalgebra of those of $\g$. We postpone such
considerations to a following section. 
\subsection{Relation of the unfolding system with a family of harmonic oscillators}
For the reasons explained in the previous section, we look for a
relationship between this system and a harmonic oscillator, and we
expect that such a relation involves the systems at hypersurfaces
of fixed negative energies. Moreover, we notice that a
reparametrization is required, since the vector field $\g$
is still not complete.\\
As stated in the previous section, we are assured that a (strictly
positive) function exists such that the vector field
$\tilde{\g}=f\g$ is complete. We look now for a reparametrization
such that the vector field $\tilde{\g}$ coincides with the
harmonic oscillator on the submanifolds of fixed negative energy.
For this purpose, we notice that  one can write $\g$ in terms of
the energy $\EL$ in the following way:
\begin{equation}
  \g=u^{\alpha}\di{y^{\alpha}}+\left(\frac{1}{2R^{4}}\EL\,y^{\alpha}-
  2\frac{u^{\beta}y^{\beta}}{R^{2}}u^{\alpha}\right)\di{u^{\alpha}}
\end{equation} so that, defining
$\Sigma_{E}:=\{(y^{\alpha},u^{\alpha})\vdash \EL=E=cost\}$, one
has:
\begin{equation}
  \g\big|_{\Sigma_{E}}=u^{\alpha}\di{y^{\alpha}}+\left(\frac{1}{2R^{4}}E\,y^{\alpha}-
  2\frac{u^{\beta}y^{\beta}}{R^{2}}u^{\alpha}\right)\di{u^{\alpha}}
\end{equation}
This form of the field will soon prove useful.\\
We have already noticed that, multiplying $\g$, a second order
vector field on $\CMcal{M}=\trq$, by a generic function
$f\in\CMcal{F}(\trq)$, the reparametrized vector field  is not
second order on $\trq$; however, in the approach of par.
\ref{sec:reptb}, it is possible to introduce on $\CMcal{M}$  a new
structure of tangent bundle with respect to which $\tilde{\g}$ is
second order. One can consider as the starting functionally
independent functions $Y^{\alpha}=y^{\alpha}$, so one has:
\begin{equation}
 \lie{\tilde{\g}}(Y^{\alpha})=f\lie{\g}(y^{\alpha})=f \,u^{\alpha}
\end{equation}
Defining $U^{\alpha}:=f \,u^{\alpha}$, one can build the following
1-1 tensor and vector field (as in Eq. \ref{eq:sdelta}):
\begin{equation}
\tilde{S}:=\de Y^{\alpha}\otimes \di{U^{\alpha}}; \qquad
\tilde{\Delta}:=U^{\alpha}\di{U^{\alpha}}
\end{equation}
so that:
\begin{equation}
  \tilde{S}(\tilde{\g})=\tilde{\Delta}
\end{equation}
that is, $\tilde{\g}$ is second order. Its \A\A second
components'' with respect to this new structure are:
\begin{equation}\label{eq:seccomprip}
  \lie{\tilde{\g}}U^{\alpha}=f\lie{\g}(f\,u^{\alpha})=f^{2}\lie{\g}u^{\alpha}+f\,u^{\alpha}\lie{\g}f
\end{equation}
These general results make possible to reformulate our search for
an $f$ such that $\tilde{\g}$ coincides with an harmonic
oscillator in the (simpler) search for an $f$ such that
$\tilde{F}^{\alpha}$ are proportional to $Y^{\alpha}$, with a term
possibly depending on the energy; more briefly:
\begin{equation}
  \tilde{F}^{\alpha}=g(\mathcal{E}_{\mathcal{L}})Y^{\alpha}
\end{equation}
For simplicity we look first for a function only of the
$y^{\alpha}$, that is $f\in\CMcal{F}(\rqq)$; so one gets:
\begin{equation}
\begin{array}{lll}
  \tilde{F}^{\alpha}&=&f^{2}F^{\alpha}+fu^{\alpha}u^{\beta}\di{y^{\beta}}f=\\
  \;&=&f^{2}\left(\cfrac{1}{R^{4}}\EL y^{\alpha}-2\cfrac{u^{\beta}y^{\beta}}{R^{2}}u^{\alpha}\right)
  +fu^{\alpha}u^{\beta}\di{y^{\beta}}f
\end{array}
\end{equation}
If one chooses\footnote{the $2$ is for convenience in
calculations, it could have been any positive real number.}
$f=2R^{2}$, the two last terms cancel each other, and the first,
when the energy is constant and negative, becomes the \A\A second
components'' of an harmonic oscillator:
\begin{equation}
  \begin{array}{lllllll}
   f&=&2R^{2}&\rightarrow&
   \tilde{F}^{\alpha}&=&2\EL\,y^{\alpha}
   -8R^{4}\cfrac{u^{\beta}y^{\beta}}{R^{2}}u^{\alpha}
   +8R^{2}u^{\beta}y^{\beta}u^{\alpha}=\\
   \,&\,&\,&\,&\,&=&2\EL\,y^{\alpha}
  \end{array}
\end{equation}
The reparametrized vector fields, in the new coordinates, reads:
\begin{equation}\label{eq:campo}
  \tilde{\g}=U^{\alpha}\di{Y^{\alpha}}+2\left(2 R^{2}u^{2}-\frac{k}{R^{2}}\right)Y^{\alpha}\di{U^{\alpha}}
\end{equation}
so, restricted to submanifolds of fixed energy
$\Sigma_{E}=\mathcal{E}_{\mathcal{L}}^{-1}(E)$, it takes the form:
\begin{equation}
  \tilde{\g}\big|_{\Sigma_{E}}=U^{\alpha}\di{Y^{\alpha}}+2E\,Y^{\alpha}\di{U^{\alpha}}
\end{equation}
It is evident that this vector field, restricted to submanifolds
of negative energy $E$, coincides with the vector field of an
harmonic oscillator of frequency $\sqrt{-2E}$:\, this was the
relationship we expected. Put in different words, for negative
energies we got a one-parameter family of harmonic oscillators,
such that the frequency of each oscillator, and so its period,
depend on this parameter, which is the energy
$\mathcal{E}_{\mathcal{L}}$ of the conformal Kepler problem in
$\rqq$; when restricted to $\Sigma_{0}$,
$\mathcal{E}_{\mathcal{L}}$ is the pull back of the energy
$\mathcal{E}_{\mathcal{L}_{K}}$ of the Kepler problem in $\rt$
(see comment after Eq.~(\ref{eq:pullbmetr})). In this way, the
orbits lying on $\Sigma_{0}$ with different
$\mathcal{E}_{\mathcal{L}}$ have different periods, depending on
$\mathcal{E}_{\mathcal{L}}$ and so on
$\mathcal{E}_{\mathcal{L}_{K}}$, thus can be related with the
orbits of the Kepler problem, a non-isocronous system; this is
what we expected taking into account the energy-period
theorem.\\
One can notice that the reparametrization involves the whole
vector field, that is the equation \ref{eq:campo} is valid
indipendently of the sign of $\EL$: so we have obtained that for
$\EL =E >0$ the reparametrized vector field coincides with a \A\A
repulsive'' harmonic oscillator (that has open orbits, in
agreement with the fact that the Kepler system has open orbits for
positive
energies), and for $\EL =0$ it coincides with the free particle.\\
As we anticipated, in this way we have obtained the completion of
the vector field $\g$: it is possible to extend the submanifold
$\Sigma_{E}$ of fixed energy to include the points corresponding
to the origin in the configuration space, the vector field
$\tilde{\g}$ being defined in these points since it coincides with
the harmonic oscillator (attractive or repulsive) on each
$\Sigma_{E}$. In this way one gets a vector field, that we still
call $\tilde{\g}$, defined on the whole
$\mathrm{T}\mathbb{R}^{3}$, and here complete (cfr \cite{mosercompl},
\cite{kummer}, \cite{iwai81}).\\
The function $f=2R^{2}$, by wich one obtains the completion, is
the very one whose existence the theorem we mentioned
assures.\\
An important comment is due about the kind of reparametrization we
have choosen. We have pointed out that the function $R^{2}$ is
\textit{one} possible function that makes possible to build the
relation between $\g$ and a family of harmonic oscillators, that
is such that:
\begin{equation}
\tilde{F}\big|_{\Sigma_{E}}
  = g(E)\, Y^{\alpha}
\end{equation}
it is by no means unique. One could have choosen, for example,
$R^{2}$ multiplied by any other (positive) function of the energy,
and so
obtained as well:\hspace{-4mm}
\begin{equation}
  \begin{array}{lllll}
   f=&2\,g(\EL )R^{2}&\rightarrow&
   \tilde{F}^{\alpha}=&2\,g(\EL )\EL\,y^{\alpha}
   -8\,g(\EL )R^{4}\cfrac{u^{\beta}y^{\beta}}{R^{2}}u^{\alpha}
   +8\,g(\EL )R^{2}u^{\beta}y^{\beta}u^{\alpha}=\\
   \,&\,\,&\,&\,&\hspace{-4mm}=2g(\EL )\EL\,y^{\alpha}
  \end{array}
\end{equation}
that is on each $\Sigma_{E}$ for $E<0$ a harmonic oscillator with
frequency $\sqrt{-2\,g(E)\,E}$.\\
So, the choice of $f=g(\mathcal{E}_{\mathcal{L}}) 2 R^{2}$ brings
just an overall multiplying factor in the expression of the
reparametrized forces (reminding equation \ref{eq:seccomprip}):
\,
clearly the reparametrized vector field coincides, also in this
case, with an harmonic oscillator (attractive or repulsive) on
each $\Sigma_{E}$, what changes is only the frequency of the
oscillator (as in the case explicitely shown above). Our choice of
$f=2\,R^{2}$ (a very immediate one) was sufficient to achieve our
purpose of relating $\g$, and so the Kepler vector field, to a
family of harmonic oscillators; one could use this arbitrariness
in the choice of a multiplying function of the energy to require
some additional properties to the reparametrized vector field.\\
Moreover, we notice that the multiplication of a vector field in
$\trq$ by $R^{2}$ correspond to the multiplication of his
projection in $\trt$ with respect to  the \ks map by $r$ (see Eq.
\ref{eq:Rr}); actually, the reparametrization of the Kepler
dynamical fields by the function $f=r$ was a common tool to
achieve its regularization (see e.g. \cite{souriau}, \cite{moser};
what they actually did was the \A\A changing of the time'' from
$t$ to $\tau$ such that $\de t=r\,\de\tau$). Gy\"{o}rgyi (in
\cite{gy}) seems to be the only one who choose a different
function $f=\mathcal{E}_{\mathcal{L}_{K}}^{-1}\, r$ (in our
notations); $\mathcal{E}_{\mathcal{L}_{K}}$ corresponds to
$\mathcal{E}_{\mathcal{L}}$ (since the latter function is its pull
back when restricted to $\Sigma_{0}$, the invariant submanifold of
our reduction procedure), so this choice is in agreement with the
ambiguity we have pointed out above.
\vspace{5mm}\\
Summarizing, we have constructed an unfolding system $\g$ for the
Kepler problem and shown that it coincides, up to
reparametrization, with a one-parameter family of harmonic
oscillators whose period depends on the energy of the Kepler
problem, such that when $\mathcal{E}_{\mathcal{L}}<0$ they are
attractive oscillators (as we anticipated on the basis of general
considerations), when $\mathcal{E}_{\mathcal{L}}>0$ they are
repulsive oscillators and the case $\mathcal{E}_{\mathcal{L}}=0$
corresponds to the free particle. For clarity we briefly recall
how the Kepler system can be obtained as a reduction of the
unfolding system $\g$. One first restricts to the submanifold
defined by the condition $R^{4}\dot{\theta}_{3}=0$ (as we
anticipated at the beginning of this section), that is invariant
for $\g$ since $\lie{\g}(R^{4}\dot{\theta_{3}})=0$. On this
submanifold the action of the group $U(1)$ given by the tangent of
the right multiplication by $\exp(i\lambda\sigma_{3})$ provides an
equivalence relation which is compatible with $\g$. In this way
one obtains the Kepler system as the reduced dynamical system.
\numberwithin{equation}{section}\section{Constants of the motion}
In the preceding sections we have constructed an unfolding system
for the Kepler problem and shown that, on submanifolds of fixed
energy, it coincides, up to reparametrization, with a harmonic
oscillator (attractive or repulsive). Using this last property, we
can obtain the constants of the motion of this system starting
with those of the harmonic oscillator. Then, we will characterize
the one parameter group we use to perform the reduction, point out
its Hamiltonian character  and then obtain the subalgebra of
symmetry for the conformal Kepler problem in 4 dimensions that
reduces to a symmetry algebra for the Kepler
problem in 3 dimensions.\\
For this purpose, we recall a known result.
\subsection{A preliminar result}\label{sec:result}
Let $G$ be the Hamiltonian symmetry group  of a given dynamical
system, and $\mathfrak{g}$ the corresponding symmetry algebra; let
us consider a one parameter Hamiltonian subgroup of $G$, whose
Hamiltonian function is $F$ and whose infinitesimal generator is
$\mathbf{Y}$. Let us reduce the dynamical system choosing the
invariant submanifold $\Sigma_{a}=\{m\in\mathcal{M}\vdash
F=a=const\}$ and the equivalence relation $\approx$ on
$\Sigma_{a}$ given by the one parameter group (which acts
preserving $\Sigma_{a}$): two points are equivalent if they are on
the same orbit of the group. With these premises, if
$\mathfrak{K}$ is a subalgebra of $\mathfrak{g}$, we want to find
under what conditions $\mathfrak{K}$ is a symmetry algebra for the
system obtained as reduction of the former by the action of this
one-parameter subgroup.
\\ First of all, the functions $f_{i}\in\mathfrak{g}$, restricted to
$\Sigma_{a}$, have to be constant on the orbits of the one
parameter group, i.e. on the integral curves of the vector field
$\mathbf{Y}$: \begin{equation}\label{eq:proiett1}
\lie{\mathbf{Y}}f=0
\end{equation}
(there is no need to explicitly restrict to $\Sigma_{a}$ since the
integral curves of $\mathbf{Y}$ belong to $\Sigma_{a}$); since
$\mathbf{Y}$ is Hamiltonian, this condition can be expressed by:
\begin{equation}\label{eq:proiett}
  \{f_{i},F\}=0
\end{equation}
This condition assures that the $f_{i}$ restricted to
\textit{each} submanifold $\Sigma_{a}$, i.e. on all the leaves of
the foliation, can be obtained as the pull-back of a function on
the quotient manifold $\tilde{\Sigma_{a}}$; if one is interested
in only one leaf, say $\Sigma_{\bar{a}}$, it is sufficient to
require:
\begin{equation}\label{eq:suff}
 \{f_{i},F\}|_{\Sigma_{\bar{a}}}=0 \quad
\mathrm{i.e.}\quad \{f_{i},F\}=c(F-\bar{a})
\end{equation}
where $c$ is a constant.
\\Let us call $\tilde{f}_{i}$ the function on the reduced space
$\tilde{\Sigma}_{a}$ that correspond to the $f_{i}$, that is
$\tilde{f}\circ\pi=f$. The condition \ref{eq:proiett1} (or
\ref{eq:proiett}) assures that the $\tilde{f}_{i}$ are constants
of the motion for the reduced system. Indeed, denoting
$\Phi_{\g}^{t}$ the flow of $\g$ on $\Sigma_{a}$ and
$\Phi_{\tilde{\g}}^{t}$ the flow of $\tilde{\g}$ on
$\tilde{\Sigma}$, it follows from the fact that $\tilde{\g}$ is
the projection of $\g$ with respect to $\pi$:
\begin{equation}
  \Phi_{\tilde{\g}}^{t}\circ\pi=\pi\circ\Phi_{\g}^{t}
\end{equation}
So one has, $\forall m\in\Sigma_{a},
\tilde{m}=\pi(m)\in\tilde{\Sigma}_{a}$:
\begin{gather}
  \tilde{f}(\Phi_{\tilde{\g}}^{t}(\tilde{m}))=\tilde{f}(\Phi_{\tilde{\g}}^{t}\circ\pi(m))=
  \tilde{f}(\pi\circ\Phi_{\g}^{t}(m))=\nonumber\\
  f\circ \Phi_{\g}^{t}(m)=f(m)=\tilde{f}(\tilde{m})
\end{gather}
i.e. $\tilde{f}$ is a constant of the motion for the reduced
system.\\
Thus, the algebra of the constants of the motion for the reduced
system is the subalgebra of the constants of the motion for the
starting system that Poisson-commutes with the Hamiltonian for the
one parameter group we use to perform the reduction.
\vspace{5mm}\\
Pursuing the program we have stated, we will first characterize
the symmetry of the unfolding system $\g$ and the one parameter
subgroup we use to perform the reduction; we will notice that we
are in the hypotesis of the previous result, which we will use to
get the constants of the motion for the Kepler problem in 3
dimensions.
\subsection{Digression: on the symmetry of the harmonic
oscillator}
In order to find the constants of the motion for the
unfolding system $\g$, we will study the ones of the
reparametrized vector field $\tilde{\g}$, since they are not
affected by reparametrization. At this aim, our main tool will be
its correspondence, on submanifolds of fixed energy, with an
harmonic oscillator, since it allows us to obtain the symmetry
group and the constants of the motion for $\tilde{\g}$ from the
ones of the harmonic oscillator.
\\
 \quad\\
We consider now an harmonic oscillator with four degrees of
freedom, defined on $\mathrm{T}\mathbb{R}^{4}$. We recall that it
admits a Lagrangian description, with Lagrangian function
$\mathcal{L}_{HO}=1/2\,(U^{\alpha}U^{\alpha}-\kappa^{2}
Y^{\alpha}Y^{\alpha})$, and corresponding Cartan 2-form:
\begin{equation}\label{eq:omegaho}
 \omega_{HO}=\de Y^{\alpha}\wedge\de U^{\alpha}
\end{equation}
and related Poisson bracket, as in Eq.~(\ref{eq:pbl}):
\begin{equation}\label{eq:pbho}
  \{f,g\}_{HO}=\frac{\partial f}{\partial U^{\alpha}}\frac{\partial g}{\partial Y^{\alpha}}
  -\frac{\partial f}{\partial Y^{\alpha}}\frac{\partial g}{\partial U^{\alpha}}
\end{equation}
Introducing complex coordinates:
\begin{equation}\label{eq:complc}
  z^{\alpha}=U^{\alpha}+i\kappa Y^{\alpha}
  \end{equation}
   the energy function
reads:
\begin{equation}
  \mathcal{E}_{\mathcal{L}_{HO}}=\frac{1}{2}\sum_{\alpha=0}^{3}z^{\alpha}\bar{z}^{\alpha}
\end{equation}
the Cartan 2-form is:
\begin{equation}
  \omega_{HO}=(2 \kappa \imath)^{-1}\mathrm{d}\bar{z}_{\alpha}
  \wedge \mathrm{d}z_{\alpha}
\end{equation}
and gives the following Poisson Bracket:
\begin{equation}\label{eq:PBhoz}
  \{f,g \}_{HO}=-2\kappa
  \imath(\partial_{\alpha}f\cdot\bar{\partial}_{\alpha}g-\bar{\partial}_{\alpha}f\cdot\partial_{\alpha}g);
  \qquad f,g \in \CMcal{F}(\trq)
\end{equation}
It follows that the quadratic functions that Poisson-commute with
the Hamiltonian  are of the form:
\begin{equation}\label{eq:costHO}
  F=(2\kappa\imath)^{-1}c_{\alpha \beta}\bar{z}^{\alpha}z^{\beta}
\end{equation}
where the $c_{\alpha\beta}$ are costant complex numbers  such that
$\bar{c}_{\alpha\beta}=-c_{\alpha\beta}$, if we require that
constants of the motion are real. The matrix we associate with
this quadratic form is a complex $4\times 4$ antihermitian matrix
$C$. Recalling that a complex antihermitian matrix can be
decomposed in the form $C=A+iB$; \: $A=-A^{T},  B=B^{T},\quad
A_{\alpha \beta},B_{\alpha \beta}\in\mathbb{R}$ with $A$ and $B$
real matrices, respectively antisymmetric and symmetric, one gets:
\begin{align}\label{eq:costanti}
  F=A_{\alpha\beta}L_{\alpha\beta}+B_{\alpha\beta}Q_{\alpha\beta}\nonumber\\
  \textrm{where}\qquad L_{\alpha \beta}&=(2\kappa)^{-1}\mathrm{Im}z^{\alpha}\bar{z}^{\beta}\nonumber \\
  Q_{\alpha \beta}&=(2\kappa)^{-1}\mathrm{Re}z^{\alpha}\bar{z}^{\beta}
  \end{align}
with $L_{\alpha \beta}=-L_{\beta\alpha }$ and $Q_{\alpha
\beta}=Q_{\beta\alpha}$; moreover, if $F= (2\kappa\imath)
^{-1}c_{\alpha \beta}\bar{z}^{\alpha}z^{\beta}$,
$G=~(2\kappa\imath)^{-1}d_{\alpha
\beta}\bar{z}^{\alpha}z^{\beta}$:
\begin{equation}\label{eq:comm}
 \{F,G\}=(2\kappa \imath)^{-1}[C,D]_{\alpha
\beta}\bar{z}^{\alpha}z^{\beta}
\end{equation}
So we have exstablished a correspondence (a Lie algebra
isomorphism) between the quadratic constants of the motion and the
complex antihermitian $4\times4$ matrices:
they form the Lie algebra $\mathfrak{u}(4)$.\\
Since one of the constants of the motion is the energy (the one
corresponding to $C=\imath \kappa \mathbb{I}$), and since it is a
central element, one usually identifies the Lie algebra
$\mathfrak{su}(4)$ as the symmetry algebra for the harmonic
oscillator. The corresponding symmetry group is $SU(4)$.
\subsection{Constants of the motion of the unfolding system}
 Since
$\tilde{\g}$, restricted to submanifold of fixed \emph{negative}
$\mathcal{E}_{\mathcal{L}}$, coincides with a harmonic oscillator
with frequency $\kappa=\sqrt{-2E}$,\, the constants of the motion
for the harmonic oscillator are constants of the motions for\,
$\tilde{\g}$\,, restricting to $\Sigma_{-}$\,,\, i.e. the part of
$\trq$ with negative $\mathcal{E}_{\mathcal{L}}$; since $\kappa$
is constant only on the submanifolds $\Sigma_{E}$, in the
extension of the costants of the motion from each $\Sigma_{E}$ to
 $\Sigma_{-}$ we have to
replace it with $\sqrt{-2\mathcal{E}_{\mathcal{L}}}$.\\
One obtains the following expressions\footnote{we have rescaled
the $Q_{\alpha\beta}$ by a factor
$\sqrt{-2\mathcal{E}_{\mathcal{L}}}$; it is a central element of
the algebra of the constants of the motion we will introduce
later} in the \A\A new'' coordinates $(Y^{\alpha},U^{\alpha})$:
\begin{align}\label{eq:costanti2}
  L_{\alpha \beta}=&
  \frac{1}{2}\left(Y_{\alpha}U_{\beta}-U_{\alpha}Y_{\beta}\right) \nonumber \\
  Q_{\alpha \beta}=& \frac{1}{2}\left(U_{\alpha}U_{\beta}-
  (2\mathcal{E}_{\mathcal{L}})Y_{\alpha}Y_{\beta}\right)
\end{align}
From this expression one can notice that $L_{\alpha\beta},
Q_{\alpha\beta}$ are well defined independently of the sign of
$\mathcal{E}_{\mathcal{L}}$, so can be extended to the whole
space, not only to the portion where
$\mathcal{E}_{\mathcal{L}}<0$. Indeed, one can check that they are
still constants of the motion for $\tilde{\g}$, whatever
the sign of $\mathcal{E}_{\mathcal{L}}$ is.\\
We point out that we  haven't yet a Poisson structure for the
constants of the motion of $\tilde{\g}$, as we got for $\g$.\, It
seems a natural choice to use the one obtained from that of the
harmonic oscillator. Since both the symplectic 2-form and the
Poisson bracket of Eqs. (\ref{eq:omegaho}), (\ref{eq:pbho}) do not
depend on the frequency of the harmonic oscillator, they can be
extended naturally from each submanifold $\Sigma_{E}$, where
$\tilde{\g}$ coincides with the oscillator of frequency
$\sqrt{-2E}$, to the whole space:
\begin{gather}
\omega_{\thicksim}=\de
Y^{\alpha}\wedge\de U^{\alpha}\nonumber\\
  \{f,g\}_{\thicksim}=\frac{\partial f}{\partial U^{\alpha}}\frac{\partial g}{\partial Y^{\alpha}}
  -\frac{\partial f}{\partial Y^{\alpha}}\frac{\partial g}{\partial U^{\alpha}}
\end{gather}
In this way, the space of the constants of the motions for
$\tilde{\g}$ becomes a Lie algebra under the Poisson bracket
\ref{eq:pbho}, which we denote $\CMcal{F}_{\tilde{\g}}$, and to
each function $f$ in $\CMcal{F}_{\tilde{\g}}$ we may associate a
the vector field $\mathbf{X}_{f}$ such that
$i_{\mathbf{X}_{f}}\omega_{\thicksim}=\de f$.
\\As we have already pointed out, the constants of the motion do
not depend on the parametrization, so $L_{\alpha\beta}$ and
$Q_{\alpha\beta}$ are constants of the motion also for $\g$. \\
We notice that, expressing $\omega_{\thicksim}$ in the \A\A old''
coordinates $(y^{\alpha},u^{\alpha})$, comparing with Eq.
\ref{eq:objects}, one has:
\begin{equation}\label{eq:omega}
  \omega_{\thicksim}=\frac{1}{2}\omega_{\mathcal{L}}
\end{equation}
i.e. they are the same, apart from a constant factor that can be
absorbed. This makes possible to build a Lie algebra isomorphism
between the algebras $\CMcal{F}_{\g}$ of the constants of the
motion for $\g$ and $\CMcal{F}_{\tilde{\g}}$ of those of
$\tilde{\g}$.\\ 
As for the expression of the Poisson bracket we got in terms of
$z$ and $\bar{z}$, if we replace $\kappa$ with $\sqrt{-2E}$ in it,
we would violate the Jacobi identity (note that
Eqs.~(\ref{eq:complc}) do not define a changing of coordinates
when $\kappa$ is replaced with $\sqrt{-2E}$). If, however, we
restrict ourselves to functions that Poisson commute with
$\mathcal{E}_{\mathcal{L}}$, we would again satisfy the Jacoby
identity on this restricted subalgebra, because on it
$\mathcal{E}_{\mathcal{L}}$ is a central element. This remark will
prove useful in a following section, where we will be allowed to
use Eqs. (\ref{eq:PBhoz}) and (\ref{eq:comm}) since we will deal
only with constants of the motion.
 \\
\qquad\vspace{3mm}\\
With these premises, we are able to investigate in a better way
the one parameter group we perform the reduction by and then get
its expression in the \A\A new'' coordinates
$(Y^{\alpha},U^{\alpha})$ and find the associated constant of the
motion. As we have already stated, this group acts as the tangent
lift (with respect to the \A\A old'' tangent bundle structure on
$\trq$) of the right multiplication by $\exp(i\lambda\sigma_{3})$:
\begin{equation}
  g \in \rqq \rightarrow  g\exp(i\lambda\sigma_{3})
\end{equation}
or, in coordinates $(y^{\alpha},u^{\alpha})$:
\begin{equation}
\left(
\begin{array}{c}
y^{1} \\
  y^{2} \\
  y^{3}\\
  y^{0}
\end{array}%
\right)\rightarrow\left(
\begin{array}{cccc}
\cos{\lambda} & -\sin{\lambda} & 0 & 0 \\
  \sin{\lambda} & \cos{\lambda} & 0 & 0\\
0 & 0 & \cos{\lambda} & -\sin{\lambda} \\
0 &  0  &  \sin{\lambda} & \cos{\lambda} \\
\end{array}
\right)\left(
\begin{array}{c}
y^{1} \\
  y^{2} \\
  y^{3}\\
  y^{0}
\end{array}%
\right)
\end{equation}
or, calling $S_{\lambda}$ the $4\times 4$ matrix above and
$\mathbf{y}^{t}$ the column vector in $\rqq$:
\begin{equation}
  \mathbf{y}^{t}\rightarrow S_{\lambda}\mathbf{y}^{t}
\end{equation}
Its infinitesimal generator is the left invariant vector field
usually denoted as $\mathbf{X}_{3}$:
\begin{equation}
  \mathbf{X}_{3}=y^{0}\di{y^{3}}-y^{3}\di{y^{0}}+y^{1}\di{y^{2}}-y^{2}\di{y^{1}}
\end{equation}
The tangent lift of the above action is given by:
\begin{equation}
 (\mathbf{y}^{t},\mathbf{u}^{t})\rightarrow(S_{\lambda}\mathbf{y}^{t},S_{\lambda}\mathbf{u}^{t})
\end{equation}
and its infinitesimal generator is the tangent lift of
$\mathbf{X}_{3}$:
\begin{gather}
  \mathbf{X}_{3}^{T}=y^{0}\di{y^{3}}-y^{3}\di{y^{0}}+y^{1}\di{y^{2}}-y^{2}\di{y^{1}}
  \qquad
  \nonumber \\ \qquad \qquad
  +u^{0}\di{u^{3}}-u^{3}\di{u^{0}}+u^{1}\di{u^{2}}-u^{2}\di{u^{1}}
\end{gather}
This vector field is Hamiltonian with respect to the 2-form
$\omega_{\mathcal{L}}$ and its Hamiltonian function is given by:
\begin{equation}
  h=i_{\mathbf{X}_{3}}\theta_{\mathcal{L}}
\end{equation}
since $\mathbf{X}_{3}^{T}$ is the tangent lift of $\mathbf{X}_{3}$
and $\omega_{\mathcal{L}}=-\de \theta_{\mathcal{L}}$, one obtains:
\begin{equation}
  h=4 R^{4}\dot{\theta}_{3}
\end{equation}
which in $(y^{\alpha},u^{\alpha})$ coordinates reads:
\begin{equation}
  h=4R^{2}(y^{0}u^{3}-y^{3}u^{0}+y^{1}u^{2}-y^{2}u^{1})
\end{equation}
If $h$ is a constant of the motion for $\g$ (as it should be, in
order to perform the reduction procedure we stated), it is a
constant of the motion also for $\tilde{\g}$. \\ 
In the \A\A natural'' coordinates for the new  tangent bundle
structure one has:
\begin{equation}
  h=2(Y^{0}U^{3}-Y^{3}U^{0}+Y^{1}U^{2}-Y^{2}U^{1})
\end{equation}
Comparing with Eq.~(\ref{eq:costanti2}), we notice that:
\begin{equation}
  h=4(L_{30}+L_{12})
\end{equation}
i.e. $h$ is the constant of the motion for $\tilde{\g}$ to which
corresponds the antisymmetric $3\times 3$ real matrix:
\begin{equation}
 N_{3}= 2\left(%
\begin{array}{cccc}
  0 & -1 & 0 & 0 \\
  1 & 0 & 0 & 0 \\
  0 & 0 & 0 & -1 \\
  0 & 0 & 1 & 0 \\
\end{array}%
\right)
\end{equation}
The vector field associated to this constant of the motion by
means of the symplectic structure $\omega_{\thicksim}$ is:
\begin{gather}
  \mathbf{X}=Y^{0}\di{Y^{3}}-Y^{3}\di{Y^{0}}+Y^{1}\di{Y^{2}}-Y^{2}\di{Y^{1}}\qquad \nonumber \\
\qquad \qquad
+U^{0}\di{U^{3}}-U^{3}\di{U^{0}}+U^{1}\di{U^{2}}-U^{2}\di{U^{1}}
\end{gather}
\vspace{8mm} \\ Summarizing, we are in the position to use the
result of section \ref{sec:result}: so, to characterize the
symmetry algebra of the reduced system, we only have to find the
constants of the motion which satisfy Eq.~(\ref{eq:suff}) with the
function $h$ Hamiltonian for $\mathbf{X}_{3}$, i.e.
$\{f_{i},h\}=c\cdot h$ (since $h=0$ on $\Sigma_{0}$), where
$c$=const.\\
For simplicity we first consider the case $c=0$, and then show
that the functions obtained in this way are all the constants of
the motion for the reduced system.
\\
For this purpose we can use the isomorphism between the Lie
algebras of the quadratic constants of the motion and of the
complex $4\times4$ antihermitian traceless matrices, previously
extablished: according to Eq.~(\ref{eq:comm}), a quadratic
function that Poisson-commutes with $h$ corresponds to a matrix
that commutes with $N_{3}$. Recalling that $C=A+iB; \: A=-A^{T},
\: B=B^{T}$, one can write the most general real traceless
symmetric (respectively antisymmetric) matrix and then impose that
its commutator with $N_{3}$ vanishes. After a lenghty but
straightforward calculation one obtains the following basis for
the Lie algebra that commutes with $N_{3}$ (cfr. \cite{iwaioa81}):
\begin{eqnarray}\label{eq:mconi}   2M_{1}=\left(%
\begin{array}{cccc}
  \; & \; & \; & -1 \\
  \; & \; & 1 & \; \\
  \; & -1 & \; & \; \\
  1 & \; & \; & \; \\
\end{array}%
\right);&\; 2M_{2}=\left(%
\begin{array}{cccc}
  \; & \; & -1 & \; \\
  \; & \; & \; & -1\\
  1 & \; & \; & \; \\
  \; & 1 & \; & \; \\
\end{array}%
\right);\nonumber \\2M_{3}=\left(%
\begin{array}{cccc}
  \; &-1 & \; & \; \\
  1 & \; & \; & \; \\
  \; & \; & \; & 1 \\
  \; & \; & -1 & \; \\
\end{array}%
\right);&\;
  2D_{1}=\left(%
\begin{array}{cccc}
  \; & \; & i\;\, & \; \\
  \; & \; & \; & i\;\, \\
  i\;\, & \; & \; & \; \\
  \; & i\;\, & \; & \; \\
\end{array}%
\right)\nonumber\\ 2D_{2}=\left(%
\begin{array}{cccc}
  \; & \; & \; & -i \\
  \; & \; & i\;\, & \; \\
  \; & i\;\, & \; & \; \\
  -i & \; & \; & \; \\
\end{array}%
\right) & \;
2D_{3}=\left(%
\begin{array}{cccc}
  i\;\, & \; & \; & \; \\
  \; & i\;\, & \; & \; \\
  \; & \;& -i  & \; \\
  \; & \; & \; & -i \\
\end{array}%
\right)
\end{eqnarray}
where the $M_{j}$ are real antisymmetric matrices and the $D_{j}$
are symmetric pure imaginary ones obtained as $D_{j}=iB_{j}$, with
$B_{j}$ real symmetric. One can check the following commutation
relations among these matrices:
\begin{flalign}
[M_{i},M_{j}]&=\epsilon_{ijk}M_{k} \nonumber\\
[D_{i},D_{j}]&=\epsilon_{ijk}M_{k}\nonumber\\
[M_{i},D_{j}]&=\epsilon_{ijk}D_{k}
\end{flalign}
These elements close a subalgebra of $\mathfrak{su}(4)$; defining
$A_{i}=\frac{1}{2}(M_{i}+D_{i})$ and
$B_{j}=\frac{1}{2}(M_{i}-D_{i})$, from their commutation
realtions:
\begin{align}
[A_{i},A_{j}]&=\epsilon_{ijk}A_{k} \nonumber\\
[B_{i},B_{j}]&=\epsilon_{ijk}B_{k}\nonumber\\
[A_{i},B_{j}]&=0
\end{align}
one recognizes the algebra
$\mathfrak{su}(2)\times\mathfrak{su}(2)$.\\
So we have obtained the result that the symmetry subalgebra that
commutes with $h$ is isomorphic to
$\mathfrak{su}(2)\times\mathfrak{su}(2)$.\\
We can now express the constants of the motion corresponding to
the $M_{j},D_{j}$ as functions on $\mathrm{T}\mathbb{R}^{4}$ using
Eqs. (\ref{eq:costHO}), (\ref{eq:costanti}), (\ref{eq:costanti2}):
\begin{gather}\label{eq:commu}
  J_{1}=\frac{1}{2}\left(Y^{1}U^{0}-Y^{0}U^{1}+Y^{3}U^{2}-Y^{2}U^{3}\right)\nonumber\\
  J_{2}=\frac{1}{2}\left(Y^{1}U^{3}-Y^{3}U^{1}+Y^{2}U^{0}-Y^{0}U^{2}\right)\nonumber\\
  J_{3}=\frac{1}{2}\left(Y^{1}U^{2}-Y^{2}U^{1}+Y^{0}U^{3}-Y^{3}U^{0}\right)\nonumber\\
  Q_{1}=\frac{1}{2}\left(U^{1}U^{3}+U^{2}U^{0}-2\mathcal{E}_{\mathcal{L}}(Y^{1}Y^{3}+Y^{2}Y^{0})\right)\\
  Q_{2}=\frac{1}{2}\left(U^{2}U^{3}-U^{1}U^{0}-2\mathcal{E}_{\mathcal{L}}(Y^{2}Y^{3}-Y^{1}Y^{0})\right)\nonumber\\
  Q_{3}=\frac{1}{4}\left[(U^{1})^{2}+(U^{2})^{2}-(U^{3})^{2}-(U^{0})^{2}-2\mathcal{E}_{\mathcal{L}}
  ((Y^{1})^{2}+(Y^{2})^{2}-(Y^{3})^{2}-(Y^{0})^{2})\right]\nonumber
\end{gather}
We notice that, since the last three constants of the motion
depend on the energy $\mathcal{E}_{\mathcal{L}}$, when we extend
them from the submanifold $\Sigma_{E}$ to the whole space, the
factor $\sqrt{-2\mathcal{E}_{\mathcal{L}}}$ is no more constant:
the Poisson commutation relations become:
\begin{flalign}
\{J_{i},J_{j}\}_{\thicksim}&=\epsilon_{ijk}J_{k} \nonumber\\
\{J_{i},Q_{j}\}_{\thicksim}&=\epsilon_{ijk}Q_{k}\nonumber\\
\{Q_{i},Q_{j}\}_{\thicksim}&=\epsilon_{ijk}J_{k}(-2\mathcal{E}_{\mathcal{L}})
\end{flalign}
Since we are interested in the subalgebra of the constants of the
motions for $\g$ that correspond to the symmetry algebra for the
Kepler problem, we now express the $J_{i}$ and $Q_{i}$ in the
coordinates $(y^{\alpha},u^{\alpha})$, since they are the natural
coordinates for the structure of tangent bundle with respect to
which $\g$ is second order:
\begin{gather}
  J_{1}=\frac{1}{4R^{2}}(y^{1}u^{0}-y^{0}u^{1}+y^{3}u^{2}-y^{2}u^{3})\nonumber\\
  J_{2}=\frac{1}{4R^{2}}(y^{1}u^{3}-y^{3}u^{1}+y^{2}u^{0}-y^{0}u^{2})\nonumber\\
  J_{3}=\frac{1}{4R^{2}}(y^{1}u^{2}-y^{2}u^{1}+y^{0}u^{3}-y^{3}u^{0})\nonumber\\
  Q_{1}=\frac{1}{8R^{4}}(u^{1}u^{3}+u^{2}u^{0})-\mathcal{E}_{\mathcal{L}}(y^{1}y^{3}+y^{2}y^{0})\nonumber\\
  Q_{2}=\frac{1}{8R^{4}}(u^{2}u^{3}-u^{1}u^{0})-\mathcal{E}_{\mathcal{L}}(y^{2}y^{3}-y^{1}y^{0})\\
  Q_{3}=\frac{1}{16R^{4}}((u^{1})^{2}+(u^{2})^{2}-(u^{3})^{2}-(u^{0})^{2})-\frac{1}{2}\mathcal{E}_{\mathcal{L}}
  ((y^{1})^{2}+(y^{2})^{2}-(y^{3})^{2}-(y^{0})^{2})\nonumber
\end{gather}
Recalling Eq.~(\ref{eq:omega}), we have to rescale these functions
 to get the same commutation relations of Eq.~(\ref{eq:commu})
with the Poisson bracket $\{\, ,
\,\}_{\mathcal{L}}$.\\
After this remark, we can recognize in the $J_{i},Q_{i}$ the
functions in $\CMcal{F}(\trq)$ corresponding to the angular
momentum and the Runge-Lenz vector in $\CMcal{F}(\trt)$
respectively. Since $\mathcal{E}_{\mathcal{L}}$ is a central
element, also in this case we can rescale the Runge-Lenz vector,
so to obtain an $\mathfrak{so}(4)$ algebra when
$\mathcal{E}_{\mathcal{L}}<0$ and an $\mathfrak{su}(2)$ when
$\mathcal{E}_{\mathcal{L}}>0$
(cfr. Eq.~(\ref{eq:marl}) and below). \\
This last consideration was sufficient to conclude that the
functions that Poisson commute with $h$ are all the constants of
the motion for the reduced system; so it is not necessary to
pursue the investigation also for values of $c$ different from
$0$.  One can arrive at the same conclusion noticing that the
above functions (actually five of them) are independent and
generate the whole space of constants of the motion for the Kepler
problem, its whole symmetry algebra.
\\The constants of the motion we have obtained coincide with those
of \cite{iwai86}, however we feel that our Lagrangian derivation
clarifies the way they arise without using any educated guess.
\numberwithin{equation}{section}\section{Conclusions} The leading
idea of this paper was that completely integrable systems should
arise from reduction of linear ones. We have considered as case of
study the Kepler problem in three dimensions. After consideration
based on general properties of the system, we have arrived at a
possible unfolding system, recovering the known relation of the
Kepler problem with a family of harmonic oscillators, in a
constuctive way, as far as it has been possible. Moreover, we have
characterized the symmetry algebra of the unfolding system and the
subalgebra of the
constants of the motion for the Kepler system.\\
It seems to us that the procedure we presented could be useful in
the unfolding of  general superintegrable systems.\\
Following the same line we hope to develop an unfolding procedure
also for quantum systems. 
\section*{Acknowledgements}
The present paper is an extended version of an invited talk given
by one of us (G.M.) at the \A\A XIX International Workshop on
Differential Geometric Methods in Theoretical Mechanics'' and
represents the main results of the laurea-thesis of the other
author (A.D.).

%
%
%
\end{document}